\documentclass[12pt,noshowpacs,nofootinbib,notitlepage,amsmath,amssymb]{revtex4-2}
\usepackage{setspace}
\allowdisplaybreaks
\usepackage{graphicx,color}
\usepackage[colorlinks=true,citecolor=blue,linkcolor=blue,urlcolor=blue]{hyperref}
\usepackage[charter]{mathdesign}
\usepackage{multirow}
\usepackage{enumerate}
\usepackage{array}
\usepackage{booktabs}

\DeclareSymbolFontAlphabet{\mathcal}{symbols}
\DeclareSymbolFont{symbols}{OMS}{xmdcmsy}{m}{n}
\DeclareSymbolFont{largesymbols}{OMX}{cmex}{m}{n}
\SetSymbolFont{symbols}{bold}{OMS}{xmdcmsy}{b}{n}

\begin{document}  
\title{\color{blue}\Large Cosmologies with turning points}

\author{Bob Holdom}
\email{bob.holdom@utoronto.ca}
\affiliation{Department of Physics, University of Toronto, Toronto, Ontario, Canada  M5S 1A7}
\begin{abstract}
We explore singularity-free and geodesically-complete cosmologies based on manifolds that are not quite Lorentzian. The metric can be either smooth everywhere or non-degenerate everywhere, but not both, depending on the coordinate system. The smooth metric gives an Einstein tensor that is first order in derivatives while the non-degenerate metric has a piecewise FLRW form. On such a manifold the universe can transition from expanding to contracting, or vice versa, with the Einstein equations satisfied everywhere and without violation of standard energy conditions. We also obtain a corresponding extension of the Kasner vacuum solutions on such manifolds.
\end{abstract}

\maketitle 

\section{The basic picture}
A fundamental premise of general relativity is that spacetime can be modelled as a 4-dimensional Lorentzian manifold. Among the defining characteristics of a Lorentzian manifold is the requirement that its metric be simultaneously smooth and non-degenerate everywhere. But the leading cosmological model based on general relativity predicts the big bang singularity, where not only the metric is degenerate, but the curvature invariants are singular. In the interest of avoiding such a curvature singularity, we shall consider relaxing the requirement that the metric be simultaneously smooth and non-degenerate. We shall make use of manifolds that are not quite Lorentzian, defined by being \textit{either} smooth everywhere \textit{or} non-degenerate everywhere, but not both simultaneously. Which of these two possibilities is realized depends on the coordinate system. Considering such manifolds from the start yields an enlarged solution space, and we shall find spatially homogeneous and isotropic solutions to the Einstein equations that avoid curvature singularities. All curvature invariants, and the components of the Einstein tensor $G_{\mu\nu}$ as well, are nonsingular for all times in both coordinate systems.

At a turning point in the evolution of the scale factor, the derivative of the scale factor vanishes. At these isolated times certain coordinate-dependent artifacts show up. When the metric at the turning point is smooth but degenerate, the degeneracy is due to the vanishing of $g_\textit{tt}$ and the inverse metric is singular. Components of the various curvature tensors with enough indices raised can then become singular even though the curvature invariants remain finite. When the metric at a turning point is non-degenerate but nonsmooth, due to a nonsmooth scale factor, all components of all curvature tensors remain finite. Instead, both the tensor components and the invariants can be be nonsmooth at the turning point. `Nonsmooth' shall always mean continuous with a noncontinuous first derivative. The Einstein equations are identically solved at all times in both coordinate systems.

The full not-quite-Lorentzian manifold may be considered to be a collection of Lorentzian manifolds, each bounded by, but not including, the time slices where the metric is either degenerate or nonsmooth. But the proper time along a timelike geodesic, between any point and a turning point, is finite, as are the corresponding coordinate times for both coordinate systems. Thus any of these Lorentzian manifolds is geodesically incomplete. Only the full not-quite-Lorentzian manifold, where $-\infty<t<\infty$, is geodesically complete. (We shall discuss the geodesics below.) This is another physically desirable feature that emerges for these manifolds. The authors of \cite{Bars:2013yba} emphasized the possible geodesic completeness of a bouncing cosmology versus its lack thereof in inflationary cosmology \cite{Borde:2001nh}. They considered evolution through the singularity itself, where the scale factor vanishes, and in this way connect a contracting phase to an expanding phase. They argued that this could happen with a Weyl-invariant matter sector. In our case the turning points occur at finite values of the scale factor, the singularities are avoided altogether, and matter can be standard.

We introduce the not-quite-Lorentzian spacetimes via the smooth metric
\begin{align}
\textit{ds}^2&=-D d'(\bar t)^2 d(\bar t)^u \textit{dt}^2+d(\bar t)( \textit{dx}^2+\textit{dy}^2+\textit{dz}^2)\label{e9}
.\end{align}
$\bar t=t/\ell$ is a dimensionless time and the nonconstant function $d(\bar t)$, its derivative $d'(\bar t)\equiv \partial d(\bar t)/\partial\bar t$, and the positive constant $D$ are also all dimensionless. The exponent $u$ is another adjustable constant. This metric turns out to generate nonsingular curvature invariants for $u>-2$ as long as $d(\bar t)>0$ for all $\bar t$. Turning points occur at $d'(\bar t)=0$ where $g_\textit{tt}$ vanishes and the metric becomes degenerate.

The violation of a basic premise of general relativity via a degenerate metric has also been considered by other authors as we detail in Section \ref{s5}. Novel to our approach is the study of metric (\ref{e9}). This metric has some surprising properties that show up when solving the Einstein equations. First, the Einstein tensor turns out to involve no more than first derivatives. And second, a solution amounts to finding the value of the exponent $u$ that is appropriate for a given equation state of a perfect-fluid matter source. After that, the function $d(\bar t)$ is still free to choose, with or without turning points.

After displaying these features of metric (\ref{e9}) in the next section, we go on to show how coordinate transformations can transform the metric into Friedmann-Lemaitre-Robertson-Walker (FLRW) form, that is where $g_\textit{tt}=-1$. In Section \ref{s2} we give two examples of new solutions that describe nonsingular bounce and oscillating universes with a normal equation of state for matter. The transformation into FLRW form yields a representation with a piecewise set of expanding and contracting FLRW cosmologies. Propagation on the spacetime is explored via geodesics and a wave equation. In Section \ref{s3} we consider the addition of spatial curvature and a more general matter content. In Section \ref{s4} we deviate from the main line of development to extend the Kasner vacuum solutions to those with turning points. We conclude with comments in Section \ref{s5}.

\section{Solutions and transformations}
We wish to solve the Einstein equations $G_{\mu\nu}= 8\pi G T_{\mu\nu}$, where $T_{\mu\nu}$ is that for a perfect fluid,
\begin{align}
T_{\mu\nu}&=(\rho+p)u_\mu u_\nu+p g_{\mu\nu}\nonumber\\
&=\textrm{diag}((-g_\textit{tt})\rho ,g_\textit{xx}p,g_\textit{yy}p,g_\textit{zz}p),
\label{e4}\end{align}
where in the last equality we have gone to the cosmic rest frame. The equation of state parameter $w$ is defined by $p=w\rho$. The nonzero components of the $G_{\mu\nu}$ for metric (\ref{e9}) are
\begin{align}
G_\textit{tt}=\frac{3 }{4 \ell^2}\frac{d'(\bar t)^2}{d(\bar t)^2}\quad G_\textit{xx}=G_\textit{yy}=G_\textit{zz}=\frac{(1 + 2u)}{4D\ell^2}\frac{1}{d(\bar t)^{1 + u}}.\label{e2}\end{align}
The components are finite and smooth everywhere, including at any turning point. There are no second derivatives and there are no differential equations to be solved. The choice $u=(3w-1)/2$ is sufficient for a solution, for then $G_{\mu\nu}= 8\pi G T_{\mu\nu}$ with (\ref{e4}) immediately gives
\begin{align}
\rho(\bar t)&=\frac{1}{8\pi G\ell^2}\frac{3}{4D}\frac{1}{d(\bar t)^{\frac{3}{2}(1+w)}},\quad p(\bar t)=w\rho(\bar t)
.\label{e8}\end{align}
Note that the particular power of $d(\bar t)$ in (\ref{e8}) is the standard result that also follows from energy conservation. We see that the Einstein equations are solved for metric (\ref{e9}) without the occurrence of the Friedmann equations and that $d(\bar t)$ is still free to choose.

Curvature invariants also have a simple, non-derivative dependence on $d(\bar t)$, as illustrated by these two, 
\begin{align}
R&=\frac{1}{\ell^2}\frac{3}{4D}(1-3w)\frac{1}{d(\bar t)^{\frac{3}{2}(1+w)}},\nonumber\\ R_{\mu\nu\rho\sigma}R^{\mu\nu\rho\sigma}&=\frac{1}{\ell^4}\frac{3}{16D^2}(9w^2 + 6w + 5)\frac{1}{d(\bar t)^{3(1+w)}}
.\label{e7}\end{align}
Thus the physical quantities in (\ref{e8}) and (\ref{e7}) and others are well behaved as long as $d(\bar t)>0$, whether or not it has turning points.

We now consider the behaviour of metric (\ref{e9}) under a coordinate transformation of the form $\bar t=f(\bar t_{\rm new})$, where the function $f$ is one-to-one and $\bar t_{\rm new}=t_{\rm new}/\ell$. Defining $d_{\rm new}(\bar t_{\rm new})=d(f(\bar t_{\rm new}))$, the new metric is
\begin{align}
\textit{ds}^2&=-D d_{\rm new}'(\bar t_{\rm new})^2 d(\bar t_{\rm new})^u \textit{dt}_{\rm new}^2+d_{\rm new}(\bar t_{\rm new})( \textit{dx}^2+\textit{dy}^2+\textit{dz}^2)
.\end{align}
The derivative is now with respect to $\bar t_{\rm new}$, and so we can say that the metric is form invariant under such transformations. Note that the factors of $\partial \bar t/\partial\bar t_{\rm new}$ in the transformation of the metric are absorbed by the transformation of the square of the derivative in the metric. If the function $d$ is one-to-one then we can go further and choose the transformation $\bar t=d^{-1}(d_{\rm new}(\bar t_{\rm new}))$. We specify that $d_{\rm new}(\bar t_{\rm new})$ is one-to-one and has the same range as $d$, but otherwise it is free to choose. If the function $d$ has turning points and is thus many-to-one, we can make a transformation as just described over each range of $\bar t$ where $d$ is one-to-one, that is, in a piecewise fashion. Combining these transformation yields a full transformation $\bar t=f(\bar t_{\rm new})$ that is one-to-one and continuous as both $\bar t$ and $\bar t_{\rm new}$ range from $-\infty$ to $\infty$.

We shall be interested in making such a coordinate transformation that takes a solution in the form of metric (\ref{e9}) with $u=(3w-1)/2$ to FLRW form. We just need to find a solution to $g_\textit{tt}=-1$, that is,
\begin{align}
D d_{\rm new}'(\bar t_{\rm new})^2 d_{\rm new}(\bar t_{\rm new})^{(3w-1)/2}=1
,\label{e10}\end{align}
subject to the constraints on $d_{\rm new}$ as we have specified. We can label such a solution by $a(\bar t_{\rm new})^2\equiv d_{\rm new}(\bar t_{\rm new})$, thus giving the metric
\begin{align}
\textit{ds}^2&=- \textit{dt}_{\rm new}^2+a(\bar t_{\rm new})^2( \textit{dx}^2+\textit{dy}^2+\textit{dz}^2)
.\end{align}
$a(\bar t_{\rm new})$ is then the standard FLRW scale factor and $a(\bar t_{\rm new})$ will satisfy the standard Friedman equations. We have transformed from a degenerate metric to a non-degenerate metric, and this requires a transformation $\bar t=f(\bar t_{\rm new})$ such that $\partial \bar t/\partial\bar t_{\rm new}\to\infty$ at each turning point. As we shall see, the result is that $a(\bar t_{\rm new})$ is nonsmooth at the turning points.

\section{Cosmology and Propagation}\label{s2}
We now consider an explicit example of a solution where $d(\bar t)$ has a turning point. We take 
\begin{align}
d(\bar t)=s+\bar t^2,\quad s>0
,\label{e13}\end{align}
for $-\infty<\bar t<\infty$. This gives a universe that is contracting for negative times, expanding for positive times, and that bounces off its minimum finite size at $\bar t=0$. The bounce is smooth and nonsingular. To go over to FLRW form we solve (\ref{e10}) with the conditions $d_{\rm new}(\pm\infty)=\infty$ and $d_{\rm new}(0)=s$. This results in 
\begin{align}
a(\bar t_{\rm new})^2&=\left(s^v+{\textstyle \frac{v}{\sqrt{D}}}|\bar t_{\rm new}|\right)^{\textstyle \frac{1}{v}},\quad v=\frac{3 w +3}{4}.
\label{e1}\end{align}
We could have started with any $d(\bar t)$ with a single turning point at $\bar t=0$ with $d(0)=s$ and $d(\pm\infty)=\infty$, since they are all solutions; and they all yield (\ref{e1}) in FLRW coordinates. The value of $\nu$ here is determined by the desired value of the equation of state parameter $w$, for example $w=1/3$ or $w=0$ for radiation or matter domination and where any $w>-1$ satisfies the null energy condition. We thus have a standard FLRW expanding phase that starts at $t_{\rm new}=0$ when the universe has finite size. This is preceded at negative times by a FLRW contracting phase.

The second example is oscillatory,
\begin{align}
d(\bar t)=s+1-\cos(\bar t),\quad s>0
.\label{e14}\end{align}
To transform to FLRW form we solve (\ref{e10}), again with $d_{\rm new}(0)=s$, but now with the constraint that $d_{\rm new}(\bar t_{\rm new})$ only increases to $2+s$ before turning back down. The result for one period of the oscillation is
\begin{align}
a(\bar t_{\rm new})^2&=\begin{cases} 
      \left(s^v+\frac{v}{\sqrt{D}}\bar t_{\rm new}\right)^{\textstyle \frac{1}{v}} & 0<\bar t_{\rm new}<{\bar t}_{\rm half} \\
      \left((2+s)^v-\frac{v}{\sqrt{D}}(\bar t_{\rm new}-{\bar t}_{\rm half})\right)^{\textstyle \frac{1}{v}} & {\bar t}_{\rm half}<\bar t_{\rm new}<2{\bar t}_{\rm half}
   \end{cases}
\end{align}
${\bar t}_{\rm half}$ is the value of ${\bar t}_{\rm new}$ at half the period,
\begin{align}
{\bar t}_{\rm half}={\frac{\sqrt{D}}{v}}\left((2+s)^v-s^v\right).
\end{align}
Each period of the original oscillation is now represented by a standard FLRW expanding phase followed by a FLRW contracting phase. The universe alternates between growing up to size $a=\sqrt{2+s}$ and then shrinking down to size $a=\sqrt{s}$. Both phases last the same amount of time, which is ${\bar t}_{\rm half}\ell$.

For both examples the expanding and contracting FLRW solutions meet at turning points where $a'(\bar t_{\rm new})$ is not continuous. Nevertheless, FLRW time is the standard cosmic time, and so when the turning point arrives, a cosmic expansion will seem to instantaneously change to a cosmic contraction, or vice versa. The acceleration $a''(\bar t_{\rm new})$ (which as usual is negative when $w>-\frac{1}{3}$) remains continuous, as does $a'(t_{\rm new})^2$. Thus the substitution of these solutions into the standard Friedman equations introduces no discontinuity, that is, the Einstein tensor $G_{\mu\nu}$ remains continuous. The matter energy density and pressure are similarly nonsmooth at turning points (unlike in (\ref{e8}) where they were smooth) in such a way that the equation of state $p(\bar t_{\rm new})=w\rho(\bar t_{\rm new})$ continues to be satisfied at all times.

We now consider the propagation of particles in these spacetimes. The particle energy is defined by $E=-u_\mu (dx^\mu/d\lambda)$ where $\lambda$ is an affine parameter in the case of massless particles, or the proper time in the case of massive particles (in which case $E$ is replaced by $E/m$). We observe the particle in the cosmic rest frame where only $u_0=-\sqrt{-g_\textit{tt}}$ is nonvanishing. If the particle travels in the $x$ direction we make use of the Killing vector $K=\partial_x$, or $K_\mu=(0,g_\textit{xx},0,0)$, and the relations
\begin{align}
K_\mu\frac{dx^\mu}{d\lambda}=\kappa,\quad g_{\mu\nu}\frac{dx^\mu}{d\lambda}\frac{dx^\nu}{d\lambda}=-\epsilon,
\end{align}
to obtain the result
\begin{align}
\frac{dt}{d\lambda}= \sqrt{\frac{\kappa^2+\epsilon g_\textit{xx}}{(-g_\textit{tt})g_\textit{xx}}}
,\end{align}
where $\epsilon=0$ or 1 and the constant $\kappa=p$ or $p/m$ for massless or massive particles respectively. This relation can be used in either coordinate system; in FLRW coordinates $t\to t_{\rm new}$.

The dependence on $g_\textit{tt}$ means that $dt/d\lambda$ diverges at a turning point for metric (\ref{e9}), while it remains finite for FLRW coordinates. But the dependence on $g_\textit{tt}$ cancels in the definition of energy, and in particular for a massless particle we have $E\propto d(\bar t)^{-\frac{1}{2}}$ and $E\propto a(\bar t_{\rm new})^{-1}$ for the two coordinate systems respectively. The difference is that $d(\bar t)$ is smooth, and so the transition between redshifting and blueshifting at a turning point is smooth for metric (\ref{e9}), while it is nonsmooth in FLRW coordinates. As we have seen, this mirrors the behaviour of other physical quantities.

The apparent coordinate velocities $dx/dt$ and $dx_{\rm new}/dt_{\rm new}$ can also be obtained from the above. For $dx/dt$, the velocity increases close to the turning point but then drops to be instantaneously zero at the turning point. $dx/dt$ is symmetric around the turning point and is nonsmooth at the turning point. For $dx_{\rm new}/dt_{\rm new}$, the velocity picks up a small nonsmooth component proportional to $|t_{\rm new}-t_{\rm new}^{\rm tp}|/\ell$ close to the turning point, where we have reinstated $\ell$. This nonsmooth behaviour can be negligible since $\ell$ is related to the size of the universe.

We may also consider the scalar wave equation $\Box \phi(\bar t,\bar X)=0$ where $X=(x,y,z)$. This equation with metric (\ref{e9}) implies that at the time of a turning point $\bar t^{\rm tp}$ we must have $\partial_{\bar t} \phi(\bar t,\bar X)|_{\bar t=\bar t^{\rm tp}}=0$, and as a result of this, the scalar quantity $g^{\mu\nu}\partial_\mu\phi\partial_\nu\phi$ remains finite. In FLRW coordinates this scalar is obviously finite, and instead, like other invariants, it is nonsmooth. In these coordinates the wave itself will have a nonsmooth component that is proportional to $|t_{\rm new}-t_{\rm new}^{\rm tp}|/\ell$ close to the turning point. Again this is typically negligible compared to the normal time variation of the wave. 

\section{Generalization}\label{s3}
Thus far we have described cosmologies where $T_{\mu\nu}$ has a single component described by $p=w \rho$, for some $w$. We shall generalize metric (\ref{e9}) in this section, so as to accommodate the more realistic situation of having several components contribute to $T_{\mu\nu}$. First let us show another need for this generalization by modifying metric (\ref{e9}) to include a spatial curvature parameterized by $k$. This is done in the usual way via spherical coordinates as follows,
\begin{align}
\textit{ds}^2&=-D d'(\bar t)^2d(\bar t)^u \textit{dt}^2+d(\bar t) \left(\frac{1}{1-k\bar r^2}dr^2+r^2 d\theta^2+r^2\sin(\theta)^2d\phi^2\right)
.\label{e3}\end{align}
Now when we take $u=(3w-1)/2$ and calculate $-G_\textit{tt}/g_\textit{tt}$ and $G_\textit{xx}/g_\textit{xx}$ we find two contributions to each. One corresponds to the original equation of state $p=w\rho$ as before. The other is proportional to $k$ and corresponds to an equation of state $p=-\frac{1}{3}\rho$, and thus $\rho(\bar t)\propto1/d(\bar t)$ for this component. This corresponds to the spatial curvature term that appears in the Einstein equations when the FLRW metric is used. In our case the Einstein equations are algebraic and they are no longer solved (that is for arbitrary $d(\bar t)$ and assuming that the actual matter has $p=w\rho$ with $w\neq-\frac{1}{3}$). In other words the additional effective contribution to $T_{\mu\nu}$ spoils the solution.

Before turning to the required generalization of metric (\ref{e9}), let us briefly consider another effective contribution to $T_{\mu\nu}$. This is due to corrections to the field equations when curvature-squared terms are added to the action. Metric (\ref{e9}) has a vanishing Weyl tensor and a vanishing Bach tensor, and so a Weyl-squared term gives no corrections. An $R^2$ term corrects the field equations with a term proportional to
\begin{align}
\left(R_{\mu\nu}-\frac{1}{4}g_{\mu\nu}R+g_{\mu\nu}\Box -\nabla_\mu\nabla_\nu\right)R
.\end{align}
By evaluating this term with metric (\ref{e9}) with $u=(3w-1)/2$, we find that it corresponds to a new effective contribution to $T_{\mu\nu}$ with
\begin{align}
\!\rho(\bar t)\propto\frac{1}{\ell^4}\frac{5 + 2\tilde w-3\tilde w^2}{d(\bar t)^{\frac{3}{2}(1+\tilde w)}},\quad \! p(\bar t)=\tilde w\!\rho(\bar t)
,\end{align}
where $\tilde w=1+2w$. This contribution has $\!\rho(\bar t)\propto 1/d(\bar t)^{3(1+w)}$ as compared to (\ref{e8}), and there is a suppression factor of order $G/\ell^2$.

So let us turn to the generalization needed to deal with a $T_{\mu\nu}$ having several components, each with its own equation of state. If we consider an effective equation of state for the complete $T_{\mu\nu}$, it is sufficient to allow this to depend on the size of the universe. We use
\begin{align}
p(\bar t)=w(d(\bar t))\rho(\bar t)
,\label{e20}\end{align}
where $p$ and $\rho$ now include all contributions. We generalize metric (\ref{e9}) to
\begin{align}
\textit{ds}^2&=-D d'(\bar t)^2F(d(\bar t)) \textit{dt}^2+d(\bar t) (\textit{dx}^2+ \textit{dy}^2+ \textit{dz}^2)
,\label{e21}\end{align}
where we now have a function $F$ instead of a power of $d(\bar t)$. It is useful to define
\begin{align}
{\cal D}(\bar t)=\exp\left({\int^{\bar t} w(d(\tilde t))\frac{d'(\tilde t)}{d(\tilde t)}\, d\!\tilde t}\right)=\exp\left({\int^{d(\bar t)} w(\tilde d)\, d\!\ln(\tilde d)}\right)
.\end{align}
${\cal D}(\bar t)\to d(\bar t)^w$ when $w(d(\bar t))$ is simply a constant $w$. We find that $G_{\mu\nu}= 8\pi G T_{\mu\nu}$ is solved with $T_{\mu\nu}$ incorporating the equation of state in (\ref{e20}) when
\begin{align}
F(d(\bar t))={\cal D}(\bar t)^\frac{3}{2}d(\bar t)^{-\frac{1}{2}}
.\label{e22}\end{align}
$G_{\mu\nu}$ is still first order in derivatives and these solutions can once again be such that $d(\bar t)$ has turning points.

The previous results in (\ref{e8}-\ref{e7}) can be updated for the generalized metric (\ref{e21}) with (\ref{e22}),
\begin{align}
\rho(\bar t)&=\frac{1}{8\pi G\ell^2}\frac{3}{4D}\frac{1}{d(\bar t)^\frac{3}{2}}\frac{1}{{\cal D}(\bar t)^\frac{3}{2}},\nonumber\\
R&=\frac{1}{\ell^2}\frac{3}{4D}(1-3w(d(\bar t)))\frac{1}{d(\bar t)^\frac{3}{2}}\frac{1}{{\cal D}(\bar t)^\frac{3}{2}},\nonumber\\ R_{\mu\nu\rho\sigma}R^{\mu\nu\rho\sigma}&=\frac{1}{\ell^4}\frac{3}{16D^2}(9w(d(\bar t))^2 + 6w(d(\bar t)) + 5)\frac{1}{d(\bar t)^{3}}\frac{1}{{\cal D}(\bar t)^{3}}
.\end{align}

The metric (\ref{e21}) continues to be form invariant under the coordinate transformations we have discussed. The transformation to FLRW form is once again of the form $\bar t=d^{-1}(a(\bar t_{\rm new})^2)$ for the finite ranges of $t$ where $d(\bar t)$ is one-to-one. Explicitly finding $a(\bar t_{\rm new})^2$ amounts to finding a solution to $g_\textit{tt}=-1$ using (\ref{e21}) and (\ref{e22}). Finding the scale factor this way is entirely equivalent to starting with the FLRW metric and using the Einstein equations to solve for the scale factor using the equation of state in (\ref{e20}).

\section{Extending Kasner}\label{s4}
Similar to the way not-quite-Lorentzian manifolds extend the solution space for homogeneous and isotropic metrics, they can also extend the solution space for metrics that are homogeneous but not isotropic. The relevant known solutions of this type are the Kasner solutions of the vacuum Einstein equations. Let us give the new solutions first; the following smooth metric solves $G_{\mu\nu}=0$ for arbitrary choice of the function $d(\bar t)$ and the constants $u_1$ and $u_2$,
\begin{align}
\textit{ds}^2&=-D d'(\bar t)^2 d(\bar t)^{2v-2} \textit{dt}^2+d(\bar t)^{-\frac{u_2u_1}{u_1 + u_2}}\textit{dx}^2+d(\bar t)^{u_1}\textit{dy}^2+d(\bar t)^{u_2}\textit{dz}^2,\label{e11}\\
&v=\frac{u_1^2 + u_2 u_1 + u_2^2}{2(u_1 + u_2)}.\nonumber
\end{align}
At turning points of $d(\bar t)$, the metric becomes degenerate.

In the special case of a power law $d(\bar t)=\bar t^\alpha$ and $D=1/\alpha^2$ we recover the original form of the Kasner metric (see eq.~(13.51) in \cite{stef}),
\begin{align}
\textit{ds}^2&=-\bar t^{2a_4} \textit{dt}^2+\bar t^{2a_1}\textit{dx}^2+\bar t^{2a_2}\textit{dy}^2+\bar t^{2a_3}\textit{dz}^2,\nonumber\\
&a_1=-\frac{\alpha}{2}\frac{u_2u_1}{u_1 + u_2},\quad a_2=\frac{\alpha}{2}u_1,\quad a_3=\frac{\alpha}{2}u_2,\quad a_4=\alpha v-1.\nonumber
\end{align}
These exponents satisfy the Kasner conditions,
\begin{align}
a_1+a_2+a_3=a_4+1,\quad a_1^2+a_2^2+a_3^2=(a_4+1)^2.
\end{align}

The quadratic curvature invariants from metric (\ref{e11}) are
\begin{align}
R_{\mu\nu\rho\sigma}R^{\mu\nu\rho\sigma}&=C_{\mu\nu\rho\sigma}C^{\mu\nu\rho\sigma}=\frac{1}{\ell^{4}}\frac{2v}{D^2}\frac{u_{1}^{2} u_{2}^{2}}{ u_{1}+u_{2}}d(\bar t)^{-4v}
.\end{align}
These curvatures remain bounded as long as $d(\bar t)$ is positive and bounded, for either sign of $v$. Since we are free to choose $d(\bar t)$ in metric (\ref{e11}) and still have a solution, we can thus extend the Kasner solutions to solutions that are everywhere nonsingular. For example we can again choose a bounce solution as in (\ref{e13}) (here we need $v>0$ since $d(\bar t)$ is unbounded from above in this case) or an oscillating solution as in (\ref{e14}).

Metric (\ref{e11}) is again form invariant under transformations of the form $\bar t=f(\bar t_{\rm new})$, such that $d(\bar t)$ is simply replaced by $d_{\rm new}(\bar t_{\rm new})=d(f(\bar t_{\rm new}))$ in the transformed metric. It is standard practice to transform the Kasner metric into a form where $g_\textit{tt}=-1$, and we can do the same for metric (\ref{e11}). The new function $d_{\rm new}(\bar t_{\rm new})$ in this case must be a solution of
\begin{align}
D d_{\rm new}'(\bar t_{\rm new})^2 d_{\rm new}(\bar t_{\rm new})^{2v-2}=1.
\end{align}
The solution is
\begin{align}
d_{\rm new}(\bar t_{\rm new})=\left(c\pm{\textstyle \frac{v}{\sqrt{D}}}\bar t_{\rm new}\right)^{\textstyle \frac{1}{v}},
\label{e15}\end{align}
for an arbitrary constant $c$, and so the transformed metric is
\begin{align}
\textit{ds}^2&=-\textit{dt}_{\rm new}^2+(c\pm{\textstyle \frac{v}{\sqrt{D}}}\bar t_{\rm new})^{2p_1}\textit{dx}^2+(c\pm{\textstyle \frac{v}{\sqrt{D}}}\bar t_{\rm new})^{2p_2}\textit{dy}^2+(c\pm{\textstyle \frac{v}{\sqrt{D}}}\bar t_{\rm new})^{2p_3}\textit{dz}^2,\\
&p_1=-\frac{u_1 u_2}{u_1+u_2}\frac{1}{2v},\quad p_2=\frac{u_1}{2v},\quad p_3=\frac{u_2}{2v}.\nonumber
\end{align}
These exponents satisfy the Kasner relations,
\begin{align}
p_1+p_2+p_3=1,\quad p_1^2+p_2^2+p_3^2=1,
\end{align}
and thus we have ended up with the common form of the Kasner metric.

The result in (\ref{e15}) assumes that the original $d(\bar t)$ is one-to-one. If it is many-to-one, then as before, the transformation must be done in a piecewise fashion over every subrange of $\bar t$ where $d(\bar t)$ is one-to-one. This will result in a set of expansions and contractions, each of the form in (\ref{e15}), that meet at turning points where the evolution, as described by the complete $d_{\rm new}(\bar t_{\rm new})$, is nonsmooth. In this section we are dealing with the vacuum Einstein equations, and these equations are identically solved by the nonsmooth metric at all times, just as they are for (\ref{e11}). Thus we have found the analog of Kasner solutions on a not-quite-Lorentzian manifold, where they can be singularity free and geodesically complete.

The Kasner solutions are such that the dependence on $t$ can be traded for any other coordinate, and the sign of each component of the metric is free to choose. This is also true for our extension of the Kasner metric, and so analogous alterations to the metric (\ref{e11}) will give solutions such as
\begin{align}
\textit{ds}^2&=-a(\bar x)^{-\frac{u_2u_1}{u_1 + u_2}}\textit{dt}^2+A a'(\bar x)^2 a(\bar x)^{2v-2} \textit{dx}^2+a(\bar x)^{u_1}\textit{dy}^2+a(\bar x)^{u_2}\textit{dz}^2.
\end{align}

\section{Comments}\label{s5}
We return to a universe that is close to being homogeneous and isotropic. Depending on how small the universe was at the last turning point, the universe could contain structure that is older than that time. Also, entropy tends to increase from one phase to the next \cite{tol}. These complications were not present in our discussion of Section \ref{s2}, where we had a single-component perfect fluid and a time evolution that was symmetric around each turning point. We generalized the equation of state parameter to be dependent on the size of the universe in Section \ref{s3}, and a further generalization would be to assume that this function of size is different for each phase, and in particular different between expanding and contracting phases. Work on the  bounce and oscillating universes has a long history; see \cite{kragh} for a historical review and \cite{Ijjas:2018qbo} for some modern motivation.

We have not discussed the stability of our solutions. But when viewed in FLRW coordinates, our solutions are locally just the standard FLRW solutions, up to a set of times of measure zero. The matter at all times satisfies standard energy conditions. It is thus difficult to see how instabilities can occur. Stability has been more of a concern for attempts to describe a nonsingular bounce cosmology on a Lorentzian manifold, where some violation of the null energy condition during the bounce seems necessary. Constructions involving nontrivial scalar field dynamics have been proposed as a way to solve that stability problem \cite{Ijjas:2016vtq}.

Something similar to our bounce cosmology, also based on a metric degenerate at $t=0$, has been studied in \cite{Klinkhamer:2019frj} (and in references therein). In that approach, the FLRW metric is modified by replacing $g_\textit{tt}=-1$ by the ansatz $g_\textit{tt}=-t^2/(t^2+b^2)$ for some constant $b$, and then obtaining the scale factor $a(t)$ via the Einstein equations that now depend on $b$. The transformation of the metric to standard FLRW form was also obtained; the new time coordinate is not continuous, jumping from a value of $-b$ to a value of $b$ as the turning point is traversed. These two times are then identified. Perturbations around the metric in the modified FLRW form were studied in \cite{Klinkhamer:2019gee} with apparently acceptable results.

Rather than stability, our solutions raise the issue of determinism. We have a metric that provides cosmological solutions to the Einstein equations for any smooth and positive $d(\bar t)$. Some such function could in principle have a set of turning points that occur randomly as a function of time, and such a cosmology would randomly switch between expansion and contraction. This could be viewed as an indeterminism in the cosmic evolution as described by the Einstein equations, when spacetime is a not-quite-Lorentzian manifold.

There may be implications for a quantum theory. It was argued in \cite{Horowitz:1990qb} that the path integral for quantum gravity should at least include all metrics with a finite action, whether or not they are degenerate at isolated points. This reference also observed what we have noted, that degeneracy at isolated points still allows the Einstein equations to be solved everywhere and it still allows curvature invariants and other scalars to be finite everywhere. The focus of that work was on degenerate metrics that could lead to topology change, but metric we have investigated does not appear to be of that type.



\begin{thebibliography}{99}
\bibitem{Bars:2013yba}
I.~Bars, P.~Steinhardt and N.~Turok,
``Local Conformal Symmetry in Physics and Cosmology,''
Phys. Rev. D \textbf{89}, 043515 (2014)
doi:10.1103/PhysRevD.89.043515
arXiv:1307.1848 [hep-th]
\bibitem{Borde:2001nh}
A.~Borde, A.~H.~Guth and A.~Vilenkin,
``Inflationary space-times are incomplete in past directions,''
Phys. Rev. Lett. \textbf{90}, 151301 (2003)
doi:10.1103/PhysRevLett.90.151301,
arXiv:gr-qc/0110012 [gr-qc].
\bibitem{stef}
Stephani, H., Kramer, D., MacCallum, M., Hoenselaers, C., Herlt, E. (2003). Exact Solutions of Einstein's Field Equations (2nd ed., Cambridge Monographs on Mathematical Physics). Cambridge: Cambridge University Press, doi:10.1017/CBO9780511535185
\bibitem{tol} R. C. Tolman, Relativity, Thermodynamics and Cosmology (Clarendon Press, Oxford, 1934).
\bibitem{kragh} Kragh, H. (2009), Continual Fascination: The Oscillating Universe in Modern Cosmology. Science in Context, 22(4), 587-612, doi:10.1017/S0269889709990172
\bibitem{Ijjas:2018qbo}
A.~Ijjas and P.~J.~Steinhardt,
``Bouncing Cosmology made simple,''
Class. Quant. Grav. \textbf{35}, no.13, 135004 (2018)
doi:10.1088/1361-6382/aac482,
arXiv:1803.01961 [astro-ph.CO]
\bibitem{Ijjas:2016vtq}
A.~Ijjas and P.~J.~Steinhardt,
``Fully stable cosmological solutions with a non-singular classical bounce,''
Phys. Lett. B \textbf{764}, 289-294 (2017)
doi:10.1016/j.physletb.2016.11.047,
arXiv:1609.01253 [gr-qc]
\bibitem{Klinkhamer:2019frj}
F.~R.~Klinkhamer and Z.~L.~Wang,
``Nonsingular bouncing cosmology from general relativity,''
Phys. Rev. D \textbf{100}, 083534 (2019)
doi:10.1103/PhysRevD.100.083534,
arXiv:1904.09961 [gr-qc].
\bibitem{Klinkhamer:2019gee}
F.~R.~Klinkhamer and Z.~L.~Wang,
``Nonsingular bouncing cosmology from general relativity: Scalar metric perturbations,''
Phys. Rev. D \textbf{101}, 064061 (2020)
doi:10.1103/PhysRevD.101.064061,
arXiv:1911.06173 [gr-qc].
\bibitem{Horowitz:1990qb}
G.~T.~Horowitz,
``Topology change in classical and quantum gravity,''
Class. Quant. Grav. \textbf{8}, 587-602 (1991)
doi:10.1088/0264-9381/8/4/007
\end{thebibliography}
\end{document}